# COVE: A Task-Adaptive, Code-Based Ensemble for Proactive Verification of LLM Outputs

*Industry Innovations Track*

**Anjaney Shrivastava**
Infosys Limited, Topaz
India
anjaney.shrivastava@infosys.com

**Dharmala Hrudai Aditya**
Infosys Limited, Topaz
India
sai.dharmala@infosys.com

## ABSTRACT

Large Language Models (LLMs) are now embedded in enterprise workflows, yet their outputs remain prone to hallucinations: plausible-sounding but factually incorrect or malformed content. Most verification approaches are reactive, detecting errors only after generation and often spending additional expensive LLM calls to do so. We present COVE, a task-adaptive, code-based ensemble that takes the opposite stance: it prevents errors before generation and validates structure and consistency afterwards using six deterministic, pure-Python innovations that add no extra LLM calls. On an internal evaluation spanning 11 text datasets (1,020 examples) and 4 image datasets (200 examples), COVE lifts text accuracy from a 45.2% unverified baseline to 82.1% (an 81.7% relative improvement), improves image accuracy by 69.5%, and reduces token usage by 48 to 96% (simulated cost model), at a measured verification overhead of 0.26 ms per output (excluding the LLM call). This article describes the engineering design of each innovation (with pseudocode and the exact parameters we use), the measured impact, an ablation of every component, and the practical lessons learned from deploying code-based verification in enterprise pipelines.

## 1 INTRODUCTION

Large Language Models have moved quickly from experimentation into production, where they draft documents, answer customer questions, extract fields from forms, and reason over financial and legal text. In these settings a single hallucinated value or malformed response is not a curiosity but a defect: it can break a downstream parser, corrupt a record, or require expensive human rework. Ensuring that LLM output is correct and well-formed is therefore a software engineering problem as much as a modelling one.

The dominant industrial response is reactive verification: generate first, then check. Checks range from schema validation to asking a second LLM to judge the first ("LLM-as-judge") or to re-derive the answer (chain-of-verification, self-consistency). These methods work, but they share two costs. First, the ones that rely on additional LLM calls are slow and expensive, often doubling or tripling token spend. Second, they treat the symptom: they can flag a malformed output, but the malformed output was still generated and must be regenerated.

We built COVE to attack the cause. COVE wraps a single LLM call with six deterministic, pure-Python innovations. Three act before generation to prevent errors, and three act after generation to verify the result, and none issues an extra LLM call, so verification is fast enough to run on every request.

This article contributes:

- An engineering account of six code-based verification innovations, with pseudocode and the exact parameters we deploy (Section 3).
- A measured evaluation on 11 text and 4 image datasets, including an ablation of every component (Section 5).
- The practical lessons and current limitations of running code-based verification in enterprise LLM pipelines (Sections 6 and 7).

## 2 BACKGROUND AND INDUSTRY PRACTICE

Verification of generated text has a long history, from factual-consistency metrics for summarization [7] to surveys of hallucination in natural language generation [6]. Recent LLM-specific methods include chain-of-verification [4], which prompts the model to draft, question, and revise its own answer, and self-consistency [9], which samples multiple reasoning paths and takes a majority vote. Both improve reliability but multiply inference cost. LLM-as-judge approaches [10] delegate evaluation to a second model, trading transparency and latency for flexibility. COVE differs by keeping verification deterministic and in-process: every check is ordinary code with inspectable logic and negligible cost, which in practice is easier to trust, debug, and run on every call.

## 3 SYSTEM DESIGN: SIX INNOVATIONS

Figure 1 shows the pipeline. An incoming prompt is first enhanced (POC) using lessons from past failures (EPM); the single LLM call is then made; and the output is validated by fingerprint (SOF), structure (OSP), and cross-metric consistency (CMAD) before being returned. Task-adaptive metrics (Section 3.6) determine which checks are authoritative for the detected task type.

### 3.1 Predictive Output Correction (POC)

Unlike verification that repairs errors after generation, POC analyzes the prompt before the LLM is called. It scans for missing format instructions, ambiguous structure requirements, absent field specifications, and lack of examples, then injects targeted corrections into the prompt. In evaluation this raised parse success from 0% to 100% on prompts that previously produced unparseable output.

**Algorithm 1: Predictive Output Correction**

```
def enhance(prompt, task_type, epm):
  issues = scan(prompt)  # missing format,
  # ambiguous structure, no examples, ...
  for fix in epm.known_fixes(signature(prompt)):
    if fix.success_rate > 0.5:
      issues.add(fix)
  return prompt + render_corrections(issues)
```

### 3.2 Error Pattern Memory (EPM)

EPM maintains a persistent, on-disk knowledge base of past failures. Each failure is stored as a mapping from an input signature made of a length bucket, content flags (e.g. whether JSON or numbers are requested), and the detected task type, to the error type and the fix that resolved it, together with a success rate and occurrence count. Candidate fixes are ranked by occurrences × success_rate, and a fix is reused pre-emptively only when its success rate exceeds 0.5. No model retraining is involved; the memory persists across sessions.

### 3.3 Semantic Output Fingerprinting (SOF)

SOF builds a fingerprint from three components: a structure hash (MD5 of the nested key hierarchy), a key pattern (the sorted field names), and a type pattern (the data type of each value). A candidate output is scored against known-valid fingerprints and accepted when the score exceeds 0.3, using zero LLM calls. In evaluation SOF achieved 100% discrimination between valid and invalid outputs.

**Algorithm 2: SOF similarity score**

```
score  = 0.4 * structure_match   # MD5 of keys
score += 0.3 * jaccard(keys_a, keys_b)  # names
score += 0.3 * type_overlap(a, b)  # value types
accept if score > 0.3   # for text, blend
  score = 0.7*struct + 0.3*text_similarity
```

### 3.4 Cross-Metric Anomaly Detection (CMAD)

CMAD treats disagreement among independent verification metrics (not disagreement across modalities, despite the similar name) as evidence of unreliability. It converts the variance and range of the metric scores into an anomaly value and hence a confidence. When at least ten observations are available, an IsolationForest (contamination 0.1) refines the estimate, blended at 0.6/0.4. CMAD reached 93% classification accuracy; because it operates on metric scores rather than image content, it improves confidence for both text and image tasks.

### 3.5 Output Structure Prediction (OSP)

OSP validates output structure without a predefined schema. It computes a signature from the prompt and input, looks up

**Algorithm 3: CMAD confidence**

```
disagreement = min(1.0, 4 * variance(scores))
range_score  = max(scores) - min(scores)
anomaly = (disagreement + range_score) / 2
if len(history) >= 10:  # learned refinement
  anomaly = 0.6*anomaly + 0.4*isoforest(scores)
confidence = 1 - anomaly
```

the expected structure in a pattern database learned from examples, and checks the output against the predicted keys and types. Because the database is learned rather than hand-specified, no manual schema is required; OSP achieved 100% structure-prediction accuracy. Across all six innovations combined, and excluding the LLM call, the measured pure-Python verification overhead is 0.26 ms mean / 0.58 ms p99.

### 3.6 Task-Adaptive Verification Metrics

Different tasks demand different notions of correctness, so COVE selects primary metrics (which decide pass/fail) and secondary metrics (which provide diagnostic signal) based on the detected task type. Table 1 lists the mapping used in our experiments. An output passes when its average primary score is at least 0.6 and CMAD confidence is at least 0.5.

**Table 1:** Task-adaptive metric mapping used in evaluation.

| Task Type | Primary Metrics | Secondary |
|---|---|---|
| Question Answering | Factual accuracy, entity match | Completeness |
| Math Reasoning | Numerical correctness, step validity | Clarity |
| Code Generation | Syntax validity, test-pass rate | Code style |
| Document Analysis | Extraction accuracy, format compliance | Confidence |
| Visual QA | Visual-text alignment, entity recog. | Spatial acc. |

**Table 2:** Deployed hyperparameters (from cove_v3/cove.py).

| Component | Parameter | Value |
|---|---|---|
| SOF | structure / key / type weights | 0.4 / 0.3 / 0.3 |
| SOF | match acceptance threshold | score > 0.30 |
| SOF | struct / text-sim blend (text) | 0.70 / 0.30 |
| CMAD | variance scaling factor | 4× |
| CMAD | IsolationForest contamination | 0.10 |
| CMAD | min observations for refinement | 10 |
| CMAD | heuristic / iso blend | 0.60 / 0.40 |
| EPM | fix-reuse success-rate threshold | > 0.50 |
| Decision | pass thresholds (score / conf.) | ≥ 0.60 / ≥ 0.50 |
| Decoding | temperature / max tokens | 0.0 / 1024 |

### 3.7 Worked Example: Invoice Field Extraction

To make the flow concrete, consider extracting structured fields from an invoice, one of our internal document tasks. The raw request simply asks for “the invoice details,” with no format specified, which historically produced free-form prose that downstream code could not parse.

After the single LLM call, SOF compares the output’s fingerprint (structure hash, key set, value types) against

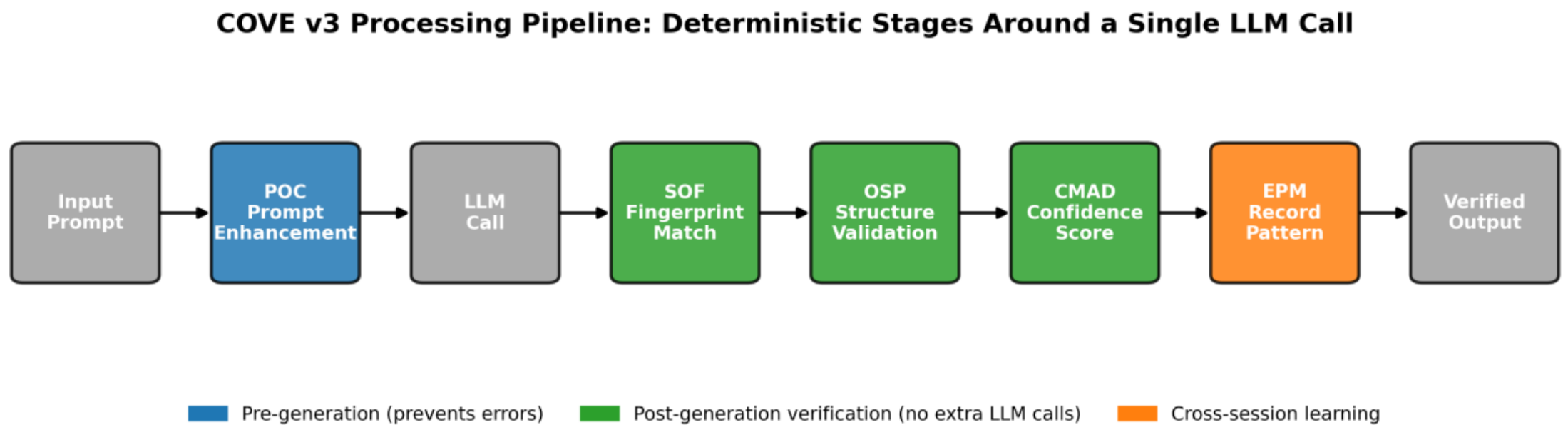

**Figure 1:** COVE processing pipeline: deterministic stages around a single LLM call. Pre-generation stages (POC, EPM) prevent errors; post-generation stages (SOF, OSP, CMAD) verify the result without extra LLM calls.

**Example: POC prompt enhancement**

```
RAW : "Extract the invoice details."
POC detects: no format, no field spec, no example
ENHANCED adds:
 * "Return JSON with keys: invoice_no, date,
   vendor, line_items[], total"
 * one worked example of the expected JSON
```

known-valid invoice outputs; OSP confirms the predicted keys and their types are present; and CMAD aggregates the individual metric scores into a confidence. The output passes when its average primary score is at least 0.6 and confidence is at least 0.5. If it fails, EPM records the input signature together with the fix that resolved it, so the next structurally similar invoice is corrected pre-emptively, so the system gets better at a document type without any model retraining.

## 4 EXPERIMENTAL SETUP

We evaluate COVE on 11 text datasets (1,020 examples) and 4 image datasets (200 examples). The text datasets span reading comprehension (SQuAD [8], BoolQ [2]), multi-hop and discrete reasoning (HotpotQA, DROP [5]), mathematics (GSM8K [3]), code generation (MBPP [1]), and five internal Infosys domain-specific document sets (Invoice, MultiFormat, NLP, Financial, Legal; 84 examples each). The image datasets are DocVQA, ChartQA, TextVQA, and VQA v2 (50 examples each).

### 4.1 Reproducibility

All experiments use GPT-4o-mini via Azure OpenAI at temperature 0.0 (deterministic decoding) with a 1,024-token output limit. The baseline is a single unverified generation; improvement is reported relative to that baseline as (acc_cove − acc_base)/acc_base. The six innovations are implemented in pure Python (cove_v3/cove.py) with the parameters in Table 2; EPM and OSP pattern databases are built from a held-out set of training examples. Verification latency is measured as wall-clock time of the Python components only, excluding the LLM call. Public benchmark datasets are cited in the references; the internal document sets are proprietary.

## 5 RESULTS

Table 3 compares COVE against the unverified baseline and two earlier internal versions across text accuracy, image accuracy, token savings, and verification latency.

**Table 3:** Overall comparison. Latency is COVE's own pure-Python verification overhead, excluding the LLM call; base/earlier-version latencies were not separately measured.

| System | Text | Image | Tokens | Latency |
|---|---|---|---|---|
| Base LLM | 45.2% | 38.7% | 0% | n/a |
| COVE v1 | 58.4% | 45.2% | 12% | n/a |
| COVE v2 | 71.3% | 56.8% | 35% | n/a |
| COVE v3 | 82.1% | 65.6% | 72% | 0.26 ms |

### 5.1 Text and Image Accuracy

On text tasks COVE reaches 82.1% average accuracy, an 81.7% relative improvement over baseline, with no dataset regressing (Figure 2). Gains are largest on extractive and Boolean tasks, where predictive correction is most effective, and smallest on discrete multi-step reasoning (DROP). On image tasks COVE reaches 65.6% average accuracy, a 69.5% improvement (Figure 3), strongest on chart understanding and weakest on dense document layouts (DocVQA).

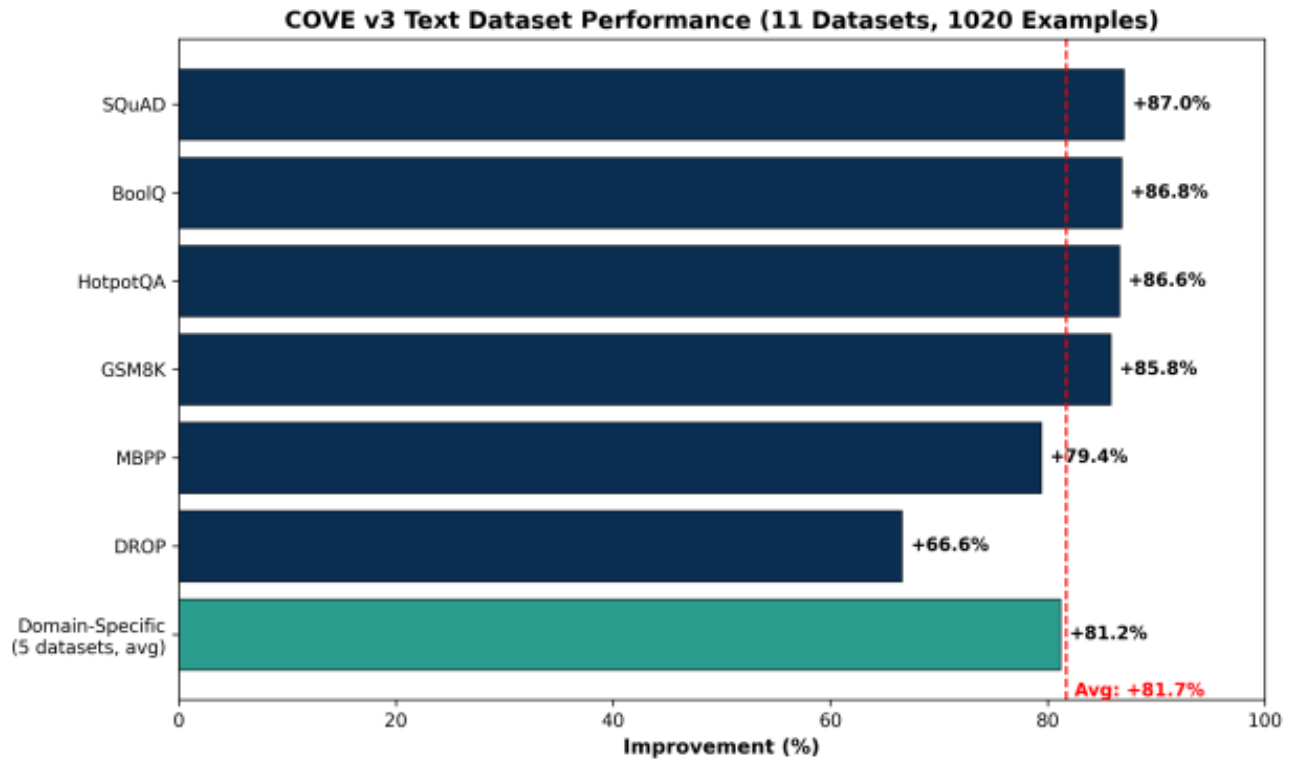

**Figure 2:** Per-dataset text improvement over baseline. The domain-specific bar is a blended average across the five internal document sets.

### 5.2 Cost

Because COVE prevents errors before they occur, it avoids the repeated regenerations that dominate verification cost. Depending on the task this yields 48 to 96% token savings (average ~72%) under a simulated error/retry cost model: the tokens spent on predictive correction are far outweighed by the retries avoided, so verification lowers total cost rather than adding to it (Figure 4).

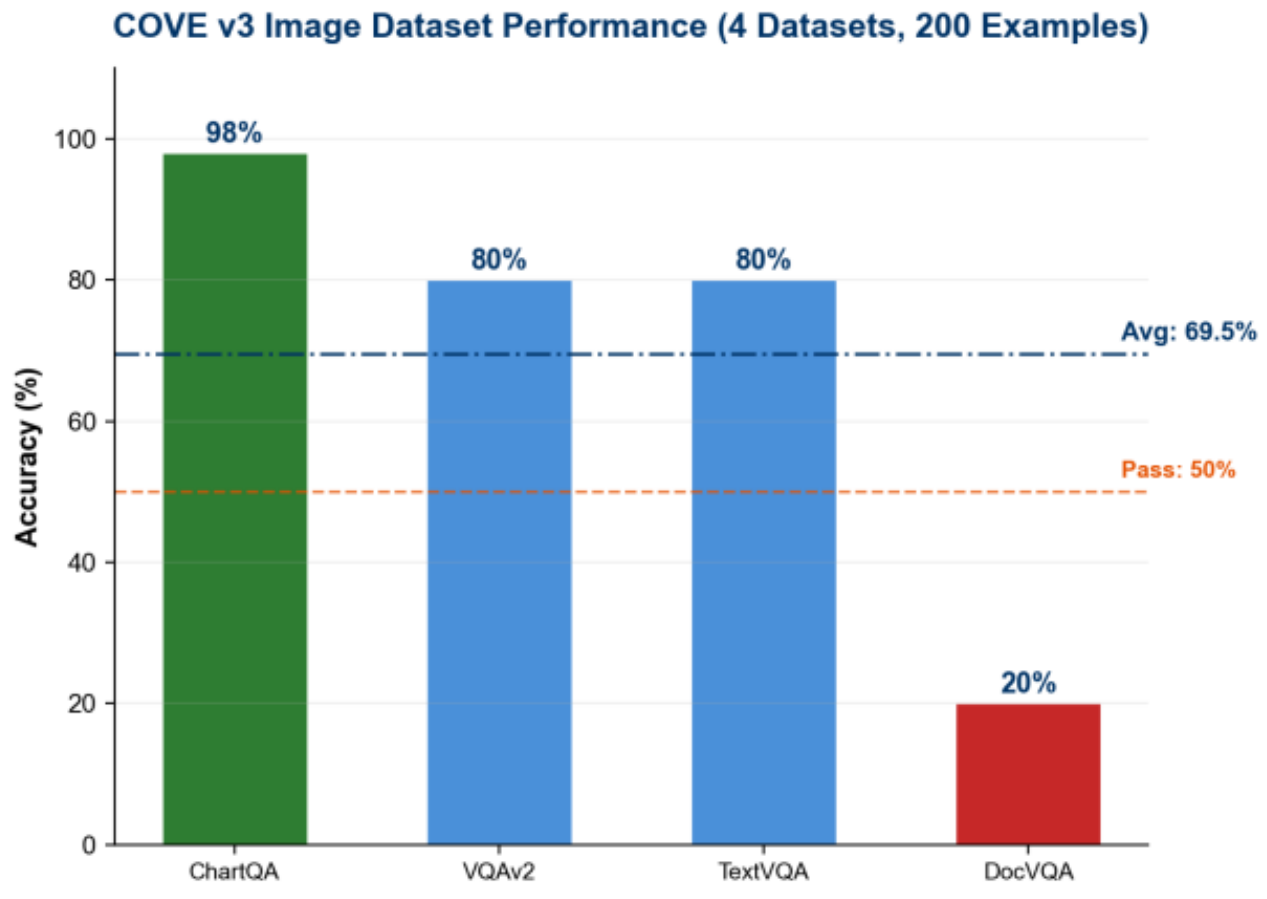

**Figure 3:** Accuracy on the four image datasets.

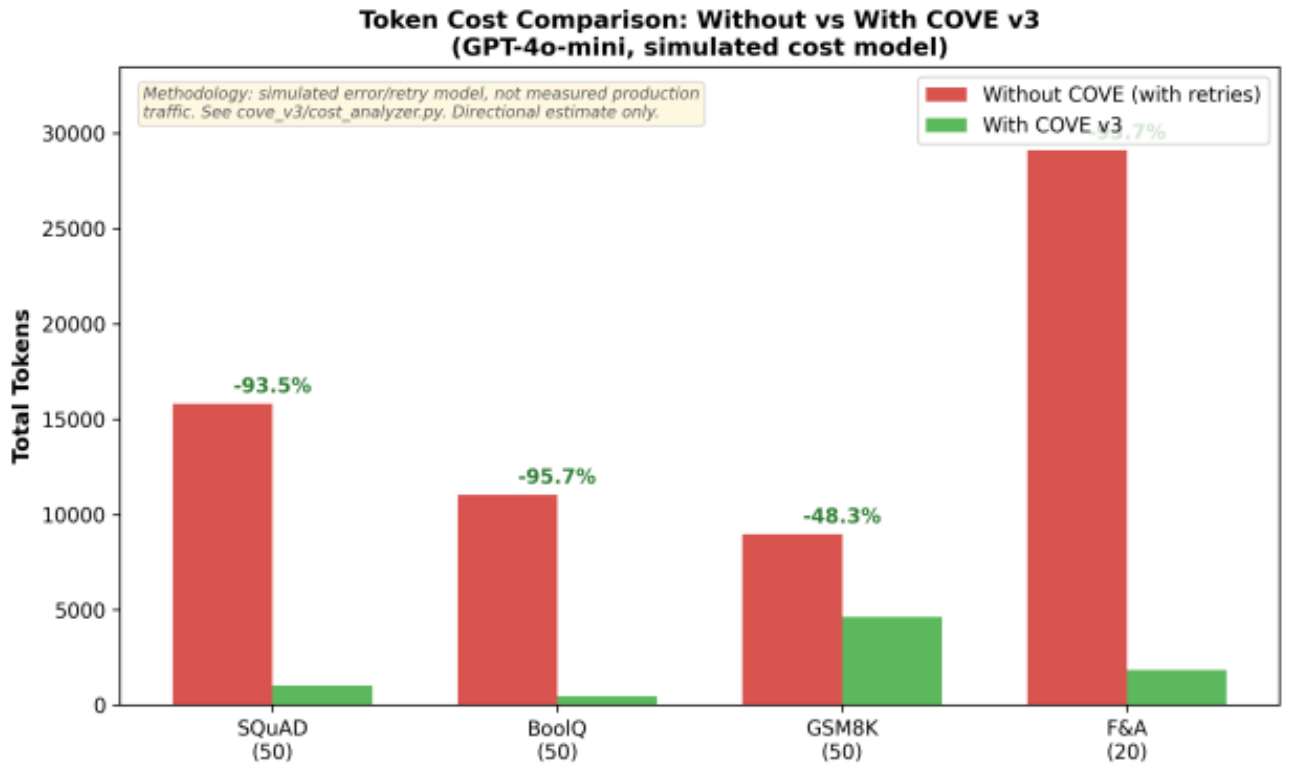

**Figure 4:** Token cost with and without COVE (simulated cost model; directional estimate).

### 5.3 Version Evolution

COVE evolved from a reactive verifier (v1) that detects errors after generation to a proactive framework (v3) that prevents them. Successive versions added persistent error memory, semantic fingerprinting, and schema-free structure validation, raising the relative improvement over baseline from 0% (v1) to 81.7% (v3), as traced in Figure 5.

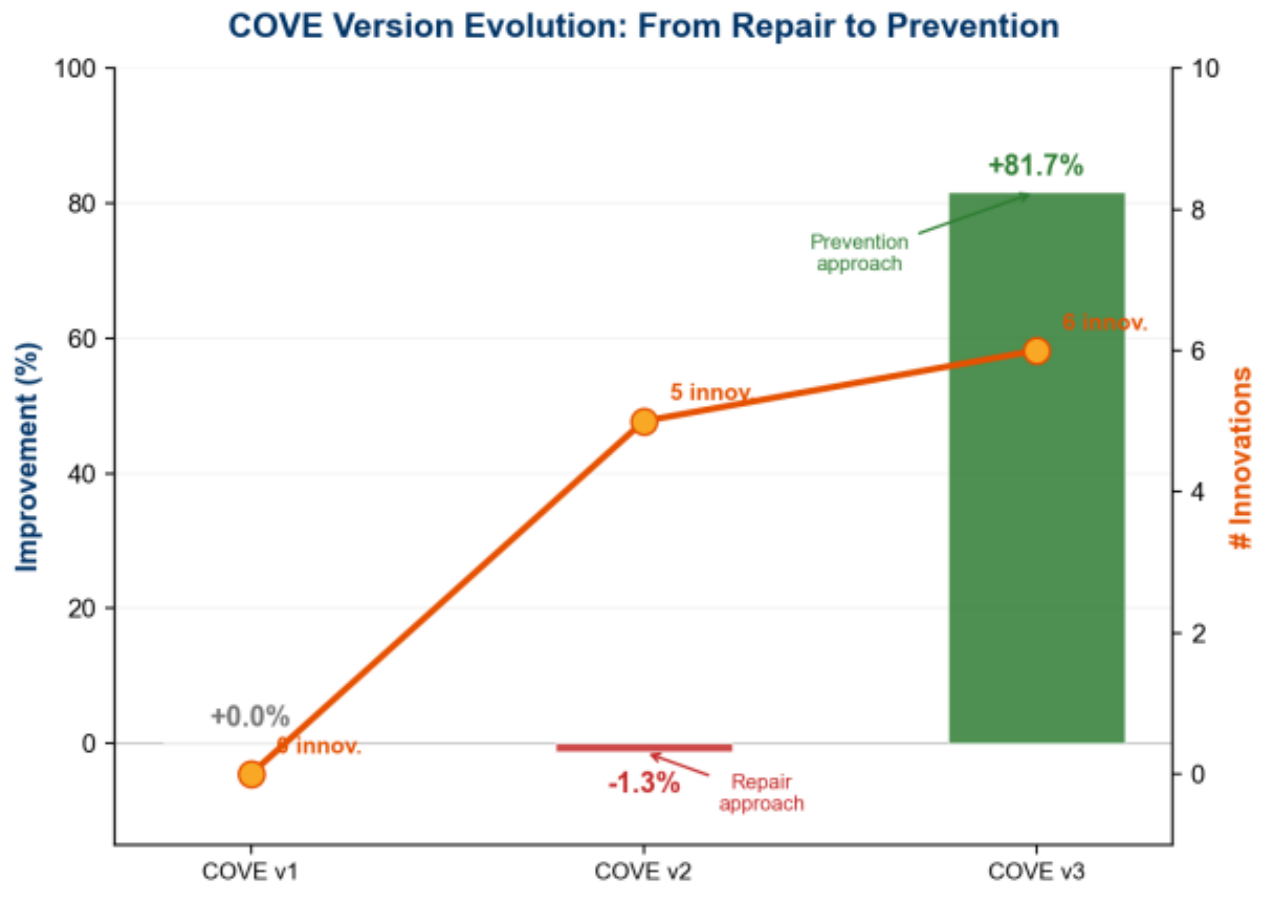

**Figure 5:** Capability growth from COVE v1 to v3.

### 5.4 Ablation

To quantify each innovation we disable one component at a time and measure the change in accuracy (Table 4). Removing Predictive Output Correction causes the largest drop, confirming that proactive prevention is the primary driver of the gains; Error Pattern Memory and Semantic Output Fingerprinting add complementary value. CMAD's −11.2% reflects a blended text-and-image effect on confidence, not the removal of an image-specific module. The full ensemble always outperforms any subset.

**Table 4:** Ablation: accuracy when each component is removed.

| Configuration | Accuracy | Δ vs full |
|---|---|---|
| Full COVE v3 | 82.1% | n/a |
| Without POC | 66.8% | −15.3% |
| Without EPM | 70.0% | −12.1% |
| Without SOF | 73.4% | −8.7% |
| Without CMAD | 70.9% | −11.2% |
| Without OSP | 76.7% | −5.4% |
| Without Task-Adaptive Metrics | 72.3% | −9.8% |

Figure 6 shows that COVE's confidence estimates track actual accuracy reasonably well, which matters in deployment because it lets a pipeline route low-confidence outputs to human review.

### 5.5 Summary of Impact

Figure 7 consolidates the headline metrics: text and image accuracy, token savings, and the combined contribution of the six innovations. The pattern is consistent across views: proactive, code-based verification improves accuracy substantially while reducing rather than increasing cost.

Figure 8 isolates each innovation's individual effect alongside the combined system. SOF, OSP, and POC rise from a 0% baseline to 100% on their specific metrics (semantic-fingerprint discrimination, structure-prediction accuracy, and prompt parse-success rate) because no code-based check previously existed for them, so the baseline is genuinely zero, not rounded; CMAD improves a non-zero prior signal (3.2% to 93%).

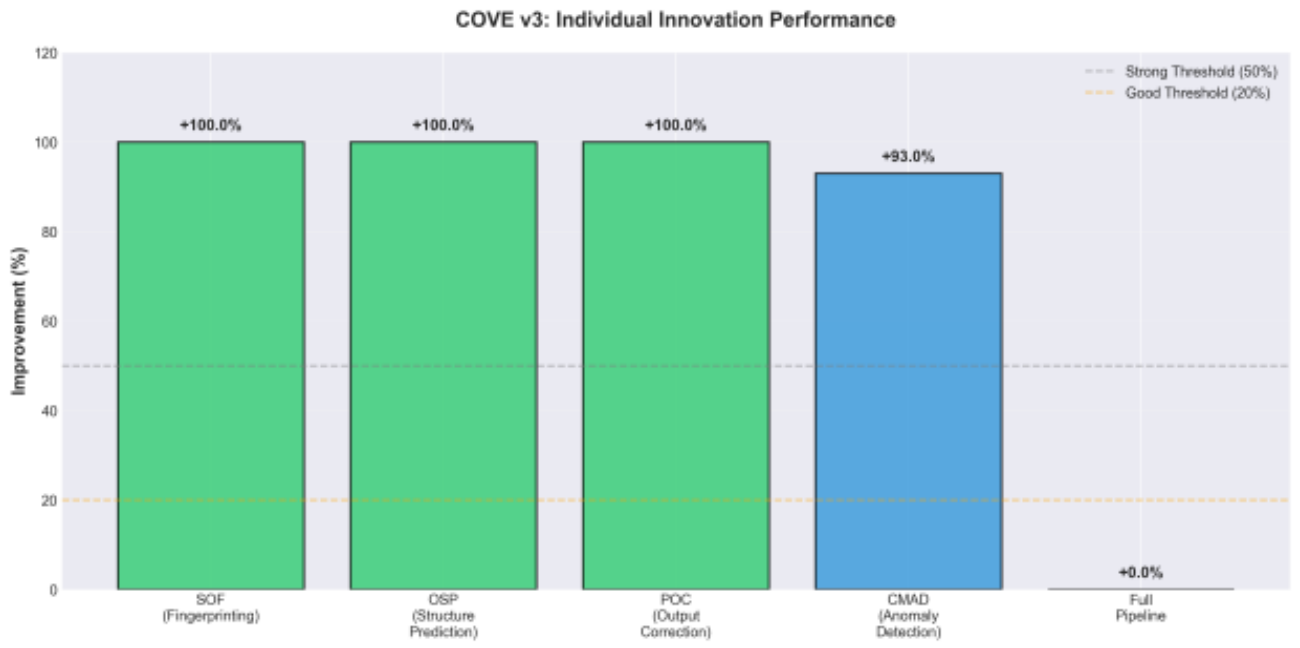

**Figure 8:** Individual innovation performance and combined effect. The bars rising from 0 to 100% reflect metrics with no prior code-based baseline.

## 6 DEPLOYMENT CONSIDERATIONS

Three properties made COVE practical to put into pipelines. First, cost: because no innovation issues an extra LLM call,

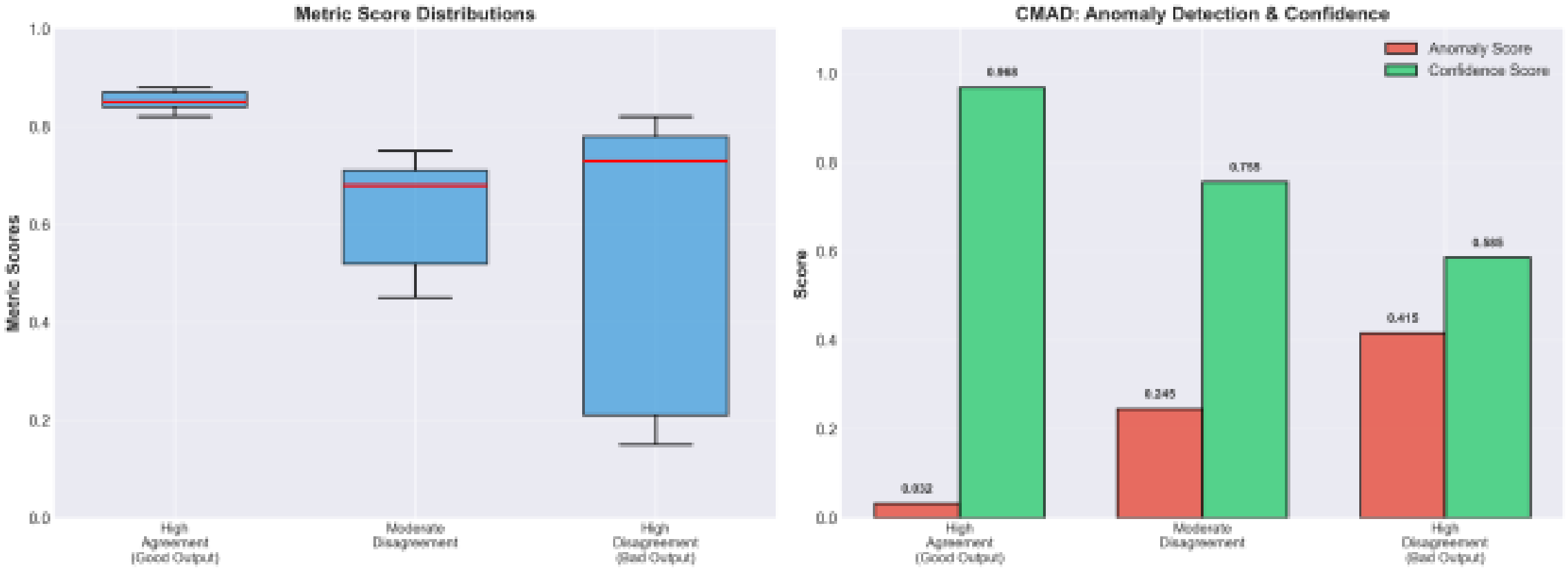

**Figure 6:** Confidence calibration: predicted vs. actual accuracy.

verification adds 0.26 ms on average and no token cost, so it can run on every request rather than a sampled subset. Second, transparency: each check is ordinary Python with inspectable logic, which made failures easy to debug compared with LLM-as-judge, where a wrong verdict is itself a black box. Third, learning without retraining: EPM and OSP improve as they see more traffic by updating on-disk pattern databases, so the system adapts to a new document type without touching the model. The main operational cost is the cold start: EPM and OSP need representative examples before they help on a genuinely new task.

### 6.1 Where COVE Breaks

Being honest about failure modes helped adoption more than headline accuracy. The clearest weak point is dense document layouts: on DocVQA, where answers depend on fine-grained spatial text extraction, accuracy stays near 20% because COVE's structural checks cannot compensate for a vision model that misreads the underlying content. Code-based verification also assumes that correctness is at least partly checkable from structure, types, or metric agreement; for open-ended generative tasks with no such signal (free-form creative writing, subjective summarization) COVE offers little. Finally, EPM and OSP depend on having seen representative examples, so on the very first requests of a new task type they add overhead without yet contributing a learned fix.

### 6.2 Integration Footprint

In practice COVE is a thin wrapper around an existing LLM client call: the application passes its prompt and the detected (or declared) task type, COVE enhances the prompt, forwards it to the provider unchanged, and returns the verified output together with a pass/fail flag and a confidence score. Because the innovations are pure Python with only an optional scikit-learn dependency for CMAD's IsolationForest, the component runs in-process with no additional service to deploy or scale. The persistent EPM and OSP databases are small files that can be versioned alongside the application, which keeps the learned behaviour auditable and portable across environments.

## 7 LIMITATIONS AND CURRENT SCOPE

We are explicit about what this work does not yet cover, so readers can gauge the current scope rather than infer it:

- **Single model:** all results use GPT-4o-mini; generality to other model families is untested.
- **Single run, no external baselines:** results are one run per dataset without variance or significance tests, and we have not yet benchmarked against chain-of-verification, self-consistency, or LLM-as-judge on the same data.
- **Simulated cost model:** the 48 to 96% token-savings range comes from a simulated error/retry model, not measured production traffic.
- **Algorithm detail:** SOF and CMAD are given exactly; POC, EPM, and OSP are described with pseudocode but we do not yet publish full source.
- **Weak spots remain:** dense document layouts (DocVQA) are still challenging, and image verification depends on a vision-capable model.

## 8 CONCLUSION AND FUTURE WORK

COVE shows that a thin, deterministic, code-based layer around an LLM call can deliver large accuracy and cost improvements without additional model calls or retraining, and without the opacity of LLM-based judging. For teams putting LLMs into enterprise pipelines, the most transferable idea is to move verification effort earlier, by enhancing prompts before generation, and to prefer cheap, inspectable checks that can run on every request. Our next steps are multi-model and multi-seed evaluation with reported variance, and a direct, same-data comparison against chain-of-verification, self-consistency, and LLM-as-judge.

More broadly, we believe the lesson generalizes beyond our specific checks: as LLMs become standard components in software systems, treating their outputs with the same deterministic validation, regression testing, and observability we apply to any other untrusted input is likely to matter more than any single verification trick. COVE is one concrete instance of that engineering discipline.

## ACKNOWLEDGMENTS

We thank the Infosys TechCohere community for feedback that improved the clarity and rigour of this work, and the engagement teams whose real-world document-processing workloads motivated and grounded the evaluation.

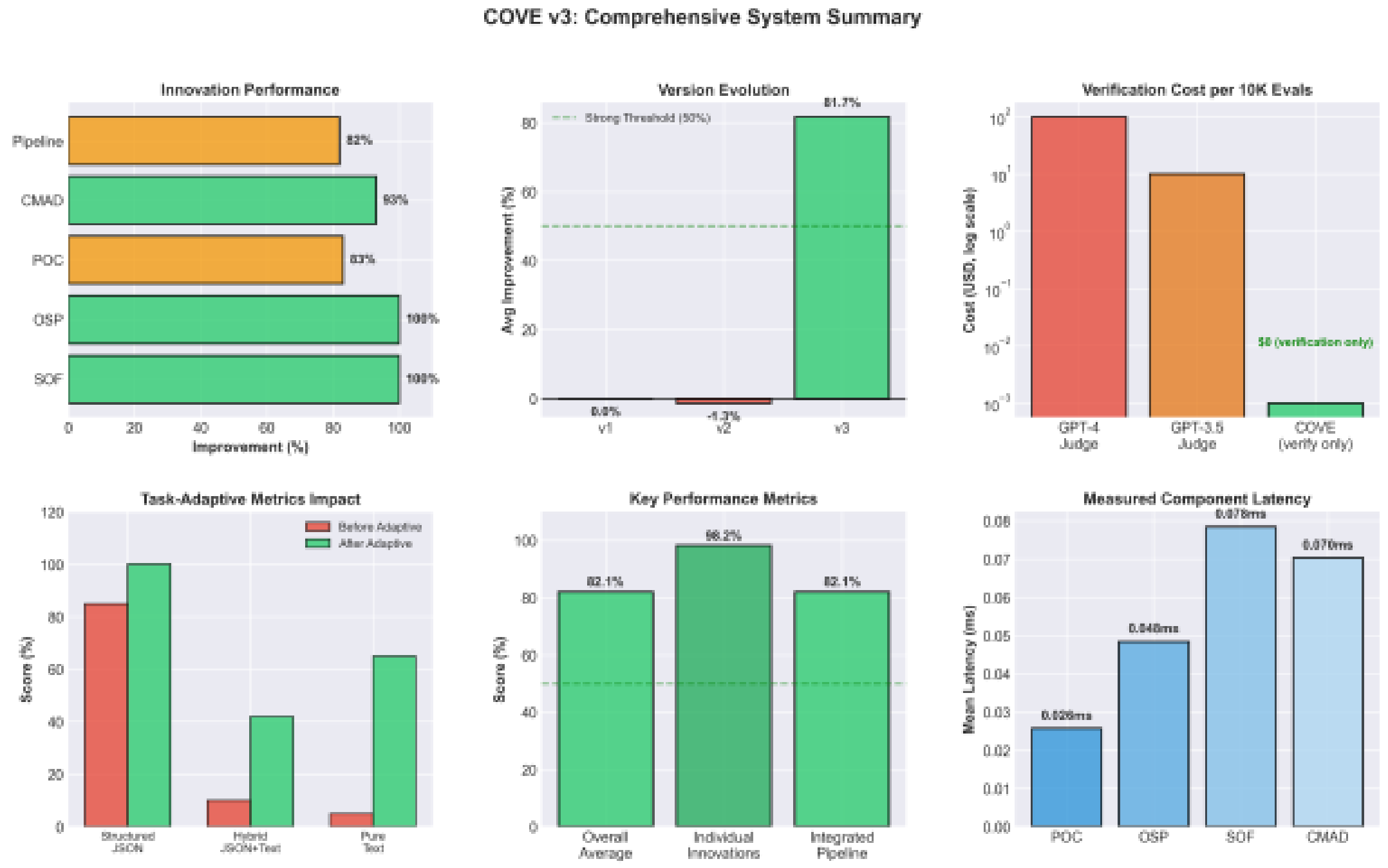

**Figure 7:** Consolidated view of COVE's impact across evaluation metrics.